\pdfoutput=1
\documentclass[12pt,preprint]{aastex}

\usepackage{graphicx}
\usepackage{natbib}
\usepackage{aas_macros}
\usepackage[section] {placeins}
\usepackage{subfigure}
\usepackage{float}
\usepackage{color}
\usepackage{amsmath}

\usepackage[scaled]{helvet}

\usepackage[T1]{fontenc}

\def\msun{{\rm\,M_\odot}}

\def\msun{{\rm\,M_\odot}}

\newcommand{\kms}{\, {\rm km\, s}^{-1}}

\def\h2{${\rm\,H_2}$}

\newcommand{\s}{\mathrm{s}}
\newcommand{\K}{\mathrm{K}}

\newcommand{\g}{\mathrm{g}}
\newcommand{\e}{\mathrm{e}}

\makeatletter 

\def\kms{{\rm\,km/s}}
\def\msun{{\rm\,M_\odot}}

\def\vol#1  {{{#1}{\rm,}\ }}

\newcount\refno
\newcount\rfno
\def\eq{$^{\the\refno\ }$\advance\refno by 1}
\def\ad{\advance\rfno by 1}

\def\clock{\count0=\time \divide\count0 by 60
     \count1=\count0 \multiply\count1 by -60 \advance\count1 by \time
     \number\count0:\ifnum\count1<10{0\number\count1}\else\number\count1\fi}

\def\myputfigure#1#2#3#4#5%
{\hskip0.03\textwidth\vskip#5pt
\makebox[0pt]{\hskip#2in
\includegraphics[width=#3\textwidth]{#1}}\vskip#4pt\hfill}

\def\newblock{\hskip .11em plus .33em minus .07em}

\newcount\refno
\newcount\rfno
\def\eq{$^{\the\refno\ }$\advance\refno by 1}
\def\ad{\advance\rfno by 1}

\definecolor{burntorange}{rgb}{1,0.4,0.2}

\definecolor{burntorange}{rgb}{1,0.4,0.2}

\usepackage{color}

\usepackage[normalem]{ulem}

\definecolor{dark-green}{RGB}{0,0,255}
\definecolor{burntorange}{rgb}{1,0.4,0.2}

\newcommand\redout{\bgroup\markoverwith{\textcolor{red}{\rule[0.5ex]{2pt}{0.8pt}}}\ULon}
\newcounter{refcounter}
\stepcounter{refcounter}

\newcounter{tempcounterb}

\newcommand{\refp}[2]{}

\begin{document}

\title{Physics of Non-Universal Larson's Relation}
 
\author{
Renyue Cen$^{1}$
} 

\footnotetext[1]{Princeton University Observatory, Princeton, NJ 08544;
 cen@astro.princeton.edu}

\begin{abstract} 

From a new perspective, we re-examine self-gravity and turbulence jointly,
in hopes of understanding the physical basis for one of the most important empirical 
relations governing clouds in the interstellar medium (ISM),
the Larson's Relation relating velocity dispersion ($\sigma_R$) to cloud size ($R$).
We report on two key new findings.
First, the correct form of the Larson's Relation is $\sigma_R=\alpha_v^{1/5}\sigma_{pc}(R/1pc)^{3/5}$,
where $\alpha_v$ is the virial parameter of clouds and $\sigma_{pc}$ is the strength of 
the turbulence, if the turbulence has the Kolmogorov spectrum.
Second, the amplitude of the Larson's Relation, $\sigma_{pc}$, is not universal, differing by 
a factor of about two between clouds on the Galactic disk and those at the Galactic center,
evidenced by observational data.

\end{abstract}
 

\section{Introduction} 

The interstellar medium (ISM) in galaxies is subject to a myriad of physical processes, including gravitational interactions,
inflow and ouflow, radiative processes, magnetic field and feedback from stellar evolution 
\citep[e.g.,][]{2007McKee}
and thus, perhaps unsurprisingly, bears a chaotic and turbulent appearance \citep[e.g.,][]{2004Elmegreen}.
The role of supersonic turbulence in interacting with the process of
gravitational collapse of molecular clouds 
has long been recognized \citep[e.g.,][]{1981Larson}.
We inquire and seek solutions as to why 
ISM clouds appear to follow a number of well defined empirical governing relations,
by examining together the two most important physical processes - turbulence and self-gravity - 
guided by a new conceptual insight.
Our goal is not set out to precisely nail down these relations, 
but rather to make sense of complex players involved, in a simple fashion, if possible.
The results we find are gratifyingly simple and accurate.

The turbulence in the ISM is driven at some large scales.
In incompressible turbulence, the structure function is derived by 
\citet[][]{1941Kolmogorov},
most notably the expression for the relation between velocity difference between two points
and their separation, $\sigma_R\propto R^{1/3}$,
based on a constant energy transmission rate through the inertia scale range.
In highly compressible turbulence, the energy transmission down through the scale
is no longer conservative, with kinetic energy also being spent to shock and/or compress the gas.
Thus, if the relation remains a scale free powerlaw,
the resulting exponent for a compressive turbulent medium is expected to be larger than $1/3$.
We will show that the right exponent is $3/5$ in this case.

Opposite to the driving scale,
the ``coherence" scale in dense cores, introduced in \citet[][]{1998Goodman},
encapsulates the transition from turbulence dominated energy regime to a subsonic regime,
where the sum of the thermal, magnetic and possibly other forms of energy dominates over turbulent energy.
The turbulence is then often thought of cascading down between these two scales.
In contrast to this simple cascading (down) of eddies in the gravity-free case,
a new conceptual notion that we put forth here 
is that the dynamic interactions between turbulence and gravity occurring on all scales
result in the formation of clouds,
within which self-gravitational force becomes important (not necessarily dominant in general),
on all scales.
While the formation of clouds is originally driven by supersonic turbulence,
gravity acts to both solidify them and in some cases detach them from the turbulence,
hence provides a feedback loop to the turbulence itself,
where the clouds may be visualized as the boundary conditions (on all scales) for the turbulence.
As such, we shall call such an additionally constrained turbulence a ``cloud bound turbulence chain" (CBTC),
as opposed to a gravity-free turbulence.
The singular coherence scale ($\sim 0.1$pc) above represents the smallest cloud of our CBTC.
Based on this conception, we attempt to rederive the (revised) Larson's Relation, 
and compare to observations.


\section{Larson's Relation: Confluence of Supersonic Turbulence and Self-Gravity}

In the ISM, self-gravity has the tendency to organize and fortify suitable regions into their own entities,
playing a countervailing role against supersonic turbulence that would otherwise produce only transient structures.
For a powerlaw radial density profile of slope $-\beta$, 
the self-gravitating potential energy is 
$W=-{3-\beta\over 5-2\beta}{GM_R^2\over R}$,
where $R$ and $M$ are the radius and mass of the cloud.
As we will show later, the density profile of gas clouds in the supersonic regime is expected to have $\beta=4/5$,
thus we will use $W=-{11\over 17}{GM^2\over R}$ for all subsequent calculations.
For a self-gravitating sphere of the same density profile, the mean velocity dispersion 
within radius $R$ is related to the 1-d velocity dispersion at separation $R$ by ${\bar \sigma_R}^2 = {11\over 14} \sigma_R^2$.
However, it proves more convenient to use $\bar \sigma_R$ instead of $\sigma_R$, since the former is a more used observable.
Hence, we shall use $\bar \sigma_R$ for all subsequent expressions; for brevity, we use $\sigma_R$ to represent $\bar \sigma_R$
hereafter.
To reduce cumbersomeness in expressions, we neglect all other forms of energy but
only to keep the gravitational energy $W$ and gas kinetic energy $K$;
it is straight-forward to include those neglected, by modifying the expression for virial parameter.
We thus define the virial parameter $\alpha_v$ as $\alpha_v=-{2K\over W}$.

The self-gravitating tendency may then be formulated as a 3-d region
in 
the four-dimensional parameter space of ($R$, $\sigma_R$, $\rho_R$, $\alpha_v$):
\begin{equation}
\sigma_R^2=\alpha_{v} {11\over 51} {G M\over R} = \alpha_v {44\pi\over 153} G \rho_R R^2 = \alpha_v {11\pi\over 51} G\Sigma_R R. 
\label{eq:grav}
\end{equation}
\noindent
where $G$ is gravitational constant, 
and $\rho_R$ and $\Sigma_R$ are the mean volume and surface density within radius $R$.
If $\alpha_v$ and $\sigma_R$ are independent, which we will show is the case,
the region would look like a thick plane.
\refp{\therefcounter}{It's important to reference Heyer's proposed modification ot Larson's Relation early on...see Figure 7 of https://iopscience-iop-org.ezp-prod1.hul.harvard.edu/article/10.1088/0004-637X/699/2/1092.}
Eq (\ref{eq:grav}) is essentially the proposed modification to Larson's Relation 
by \citet[][]{2009Heyer}. More comparisons will be made in \S 3.

\citet[][]{1941Kolmogorov} power spectrum is derived 
for homogeneous and isotropic three-dimensional subsonic turbulence in incompressible flows,
valid in the energy conserving inertial range.
In contrast, the kinetic energy in the supersonic compressible turbulence in ISM is dissipative on all scales due to shocks
and radiative processes in the ISM.
It thus, at first instant, might suggest that
the Kolmogorov turbulence may provide an inadequate description of the compressible turbulence of the ISM.

\citet[][]{1983Fleck} suggests 
that the relation between a scaled velocity $v_R$ and scale $R$ of compressible turbulence be expressed as  
\begin{equation}
\label{eq:kol}
\begin{split}
v_R \equiv \rho_R^{1/3} \sigma_R =A R^{1/3},
\end{split}
\end{equation}
\noindent
which constitutes a plane in the parameter space of ($R$, $\sigma_R$, $\rho_R$),
generally different from that of self-gravity (Eq \ref{eq:grav}),
where $A$ is a constant.
The expression essentially asserts that 
a constant volumetric energy density transfer rate in compressible flow is transmitted down the turbulence cascading scale.
Eq (\ref{eq:kol}) reduces to the original Kolmogorov form for incompressible flow that is a line in the two-dimensional parameter space
of ($R$, $u_R$).
A formal proof of the existence of an inertial range for highly compressible turbulence 
is given by \citet[][]{2011Aluie,2013Aluie}, validating the density-weighted velocity formulation.
Importantly, numerical simulations
show that the spectrum of $v_R$ indeed follows remarkably well the Kolmogorov spectrum
for the isothermal ISM \citep[e.g.,][]{2007Kritsuk, 2013Kritsuk}.
We thus continue to use the nomenclature of Kolmogorov compressible turbulence, despite its oxymoronic sounding,
given the spectral slope we adopt and its empirical validity to describe the turbulence of the isothermal ISM.
The general physical arguments and quantitative conclusions reached
are little altered with relatively small variations of the slope of the turbulence power spectrum.

As a related note, in the subsonic compressible turbulence, with gravity also playing
an important role, such as in dark cores in molecular clouds,
the physical premise for the argument of energy transmission through the inertia scale range ceases to apply
with respect to the total velocity.
This may be understood in that the turbulence chain driven at some large scales
no longer is the primary driver of velocity in the subsonic regime.
Rather, the velocity field is driven jointly by turbulence, thermal (and possible other forms of) pressure, 
and gravity \citep{1983Myers}.

Combining Eq (\ref{eq:grav}) and Eq (\ref{eq:kol}) gives
\begin{equation}
\label{eq:sigmaR}
\begin{split}
\sigma_R = \alpha_v^{1/5} ({44\over 153} A^3 G)^{1/5} R^{3/5}.
\end{split}
\end{equation}
\noindent
Because $A$ is unknown but a constant, we simply introduce another parameter, $\sigma_{pc}$, 
which denotes the 1-d mean turbulence velocity dispersion 
within a region of radius $1$ parsec,
to express the strength of the turbulence.
Now Eq (\ref{eq:sigmaR}) is simplified to 
\begin{equation}
\label{eq:sigmaR2}
\begin{split}
\sigma_R = \alpha_v^{1/5} \sigma_{pc} ({R\over 1pc})^{3/5}.
\end{split}
\end{equation}
\noindent
Looking at 
Eq (\ref{eq:sigmaR2}),
it may seem puzzling as to why the virial parameter $\alpha_v$ appears in this expression that 
is supposedly an expression of the strength of the turbulence chain.
But it is expected.
The appearance of $\alpha_v$ (and the disappearance of gas density $\rho$) in this expression
reflects the feedback of the boundary condition at the clouds 
that terminates the turbulence chain at the small scale end, in lieu of gas density.
To see that we may express the cloud density in terms of $\sigma_{pc}$:
\begin{equation}
\label{eq:npc}
\begin{split}
n_{pc} = \alpha_v^{-3/5} {153\over 44\pi} {\sigma_{pc}^2\over G m_p (1pc)^2} = 1.04\times 10^4 cm^{-3}\alpha_v^{-3/5} ({\sigma_{pc}\over 1\kms})^2, 
\end{split}
\end{equation}
\noindent
where $n_{pc}$ is the mean density within a cloud of radius $1$pc with a virial parameter $\alpha_v$.

\begin{figure}[h!]
\centering
\vskip -0.0cm
\resizebox{4.5in}{!}{\includegraphics[angle=0]{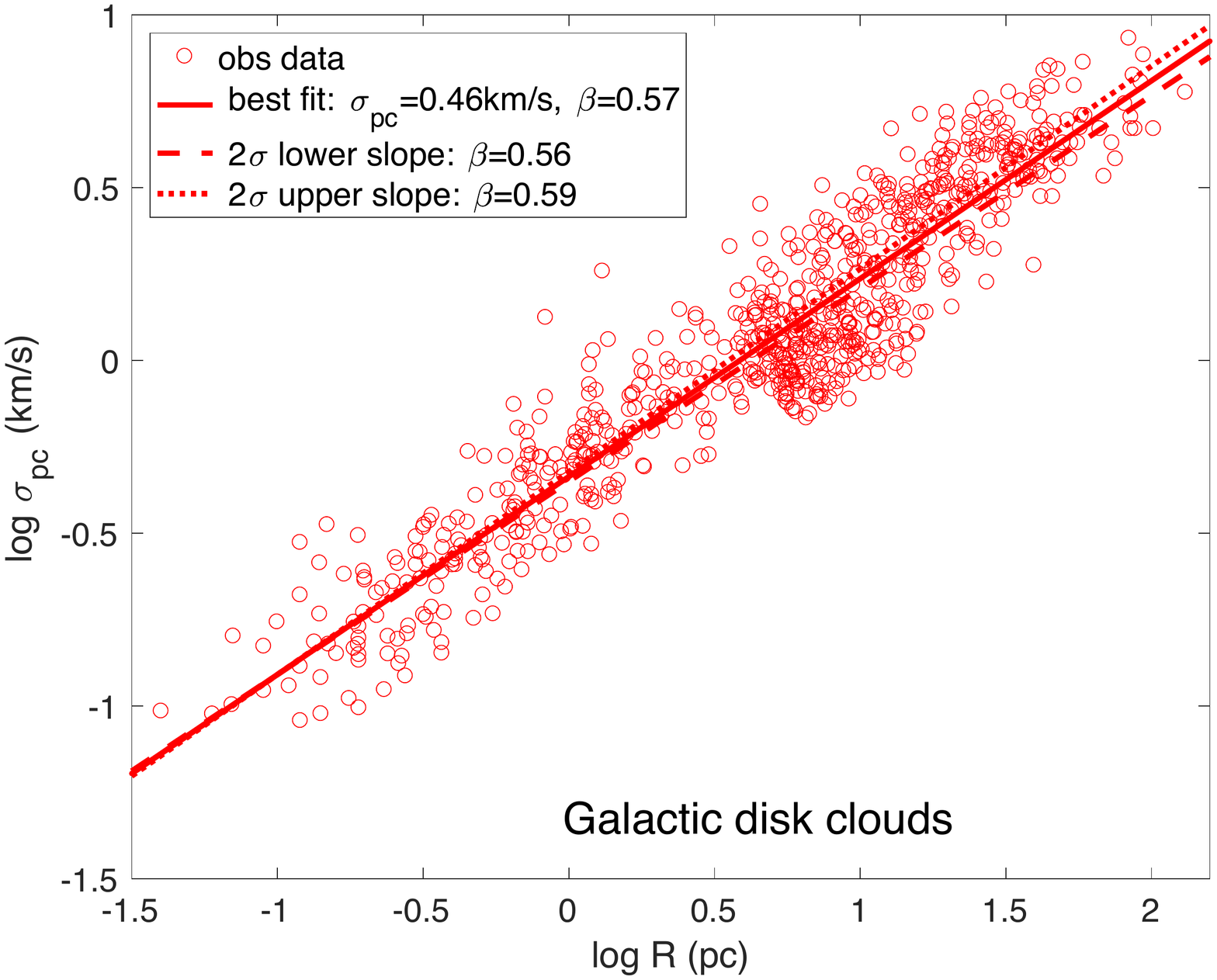}}
\resizebox{4.5in}{!}{\includegraphics[angle=0]{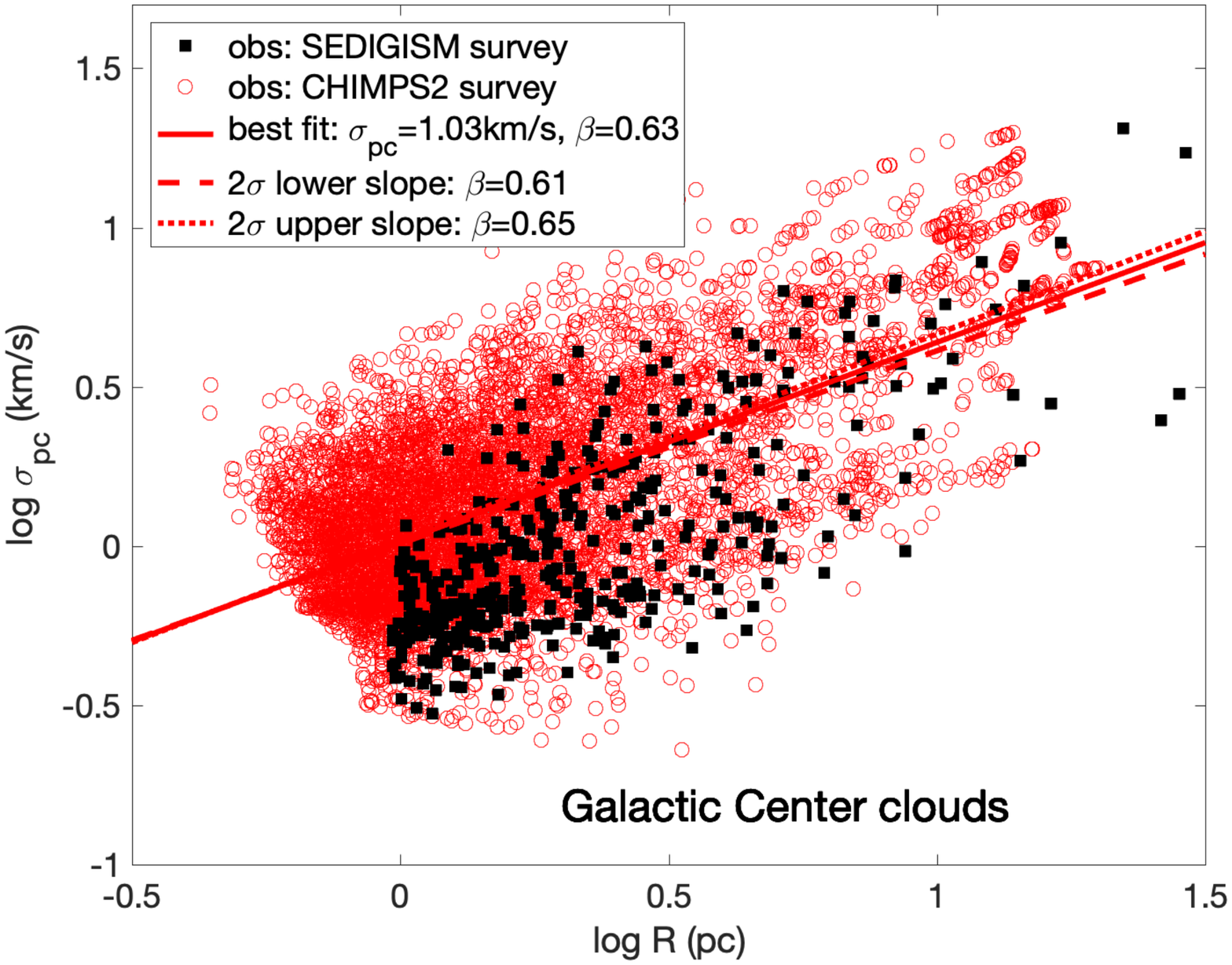}}
\vskip -0.5cm
\caption{
Top panel shows the velocity as a function of its size
for the observed molecular clouds on the Galactic disk (open red circles),
from \citet{1986Dame}, \citet{1987Solomon},\citet{2001Heyer},
\citet{2004Heyer},  
\citet{2006Ridge}, 
\citet{2008Narayanan}  
and
\citet{2013Ripple}.  
Bottom panel shows the velocity as a function of its size,
for the observed molecular clouds at the Galactic center 
from the CHIMPS2 survey \citep[][]{2020Eden} (open red circles) and 
the SEDIGISM \citep[][]{2020DuarteCabral} (solid black squares). 
In each panel, we show as red solid line as the best powerlaw fit using linear regression,
along with
the $2\sigma$ upper and lower slopes shown as dotted and dashed lines, respectively,
obtained with bootstrapping.
}
\label{fig:sigR}
\end{figure}

Eq (\ref{eq:sigmaR2}) is the (revised) Larson's first relation, relating 
the velocity dispersion to the size of the cloud.
Let us now proceed to compare this relation to observational data.
Figure (\ref{fig:sigR}) shows the observational data along with best powerlaw fits.
We fit the data to a powerlaw of the form 
\begin{equation}
\label{eq:sigmaR3}
\begin{split}
\sigma_R = \alpha_v^{1/5} \sigma_{pc} ({R\over 1pc})^\beta,
\end{split}
\end{equation}
leaving both the amplitude $\sigma_{pc}$ and the exponent $\beta$ as two free parameters.
Moreover, we perform bootstrap resampling to obtain upper and lower $2\sigma$ limits of the fitting parameters
by fitting both parameters.
We find the best parameters and the $\pm 2\sigma$ limits for the disk clouds to be 
\begin{equation}
\begin{split}
{\rm best\ fit:}\quad \sigma_{pc} &= 0.46\pm 0.03\kms \quad\quad {\rm and} \quad\quad \beta=0.57\pm 0.02 \\
{\rm +2\sigma:}\quad \sigma_{pc} &= 0.48 \quad\quad\quad\quad\quad\quad\quad\ \ {\rm and} \quad\quad \beta=0.59 \\
{\rm -2\sigma:}\quad \sigma_{pc} &= 0.45 \quad\quad\quad\quad\quad\quad\quad\ \ {\rm and} \quad\quad \beta=0.56,
\end{split}
\label{eq:disk}
\end{equation}
\noindent
shown as the solid, dotted and dash lines, respectively, 
in the top panel of Figure (\ref{fig:sigR}).
Repeating the calculation for the clouds at the Galactic center yields
the best parameters and the $\pm 2\sigma$ limits
\begin{equation}
\begin{split}
{\rm best\ fit:}\quad \sigma_{pc} &= 1.03\pm 0.01\kms \quad\quad {\rm and} \quad\quad \beta=0.63\pm 0.01 \\
{\rm +2\sigma:}\quad \sigma_{pc} &= 1.02 \quad\quad\quad\quad\quad\quad\quad\ \ {\rm and} \quad\quad \beta=0.65 \\
{\rm -2\sigma:}\quad \sigma_{pc} &= 1.05 \quad\quad\quad\quad\quad\quad\quad\ \ {\rm and} \quad\quad \beta=0.61,
\end{split}
\label{eq:center}
\end{equation}
\noindent
shown as the solid, dotted and dash lines, respectively, 
in the bottom panel of Figure (\ref{fig:sigR}).
We note that the errorbars of the best using the linear regression method 
is not necessarily consistent with and often larger than the $2\sigma$ range obtained using bootstrap,
due to the latter's larger sample size with bootstrapping.
The discrepancy is more noticeable for the disk clouds due to the smaller observational data sample size,
as compared to that of the Galactic center clouds.
Nevertheless, even in the absence of this shift for the best slope,
the traditional exponent of the Larson's Relation of $1/2$ is inconsistent with 
the disk data at 100\% level if bootstrap is used and at $3.5\sigma$ if the direct regression is used,
whereas a slope of $0.6$ is about $1.5\sigma$ away.
If considering the clouds at the Galactic center, the contrast is still larger.

So far, we have not considered possible (perhaps different) systematics for the observations
of the Galactic disk clouds as compared to the Galactic center clouds.
The fact that the best fitting slope of the disk clouds of $0.57$ and that of the Galactic center clouds of $0.63$
equidistantly flank our proposed slope of $0.60$ is intriguing.
It may be caused by some additional physics that are not considered in our simplified 
treatment but operates to varying degrees of importance in the cases.
It may also be caused by data inhomogeneities in the plotted plane, which may already be visible.
We shall take the simpler interpretation that both slopes are intrinsically equal to $0.60$
and the apparent values are due to some observational systematics,
although we are not in a position to justify this assertion.

This new Larson's Relation with the exponent $3/5$ is in excellent agreement with observational data.  
\citet[][]{2004Heyer} measure
the value of the scaling exponent of $0.59\pm 0.07$ in the spatial range of $1-50$pc [corresponding to
the original range of \citet[][]{1981Larson} and \citet[][]{1987Solomon}],
while fitting the entire spatial range of $0.03-50$pc probed they get $0.62\pm 0.09$.



It is clear now that it is not just the gravity alone that gives rise to the Larson's Relation,
rather it is a combination of gravity and turbulence physics that naturally yields it.
Larson 1981 invoked virial equilibrium to explain his relation.
What is new here is that the intersection of gravity and turbulence provides a significantly
better fit for data. 
Forcing the slope of $0.6$ to both data sets, the best fit $\sigma_{pc}$ is found to be
\begin{equation}
\begin{split}
\sigma_{pc} &= 0.44\pm 0.02 \kms \quad\quad {\rm for\ Galactic\ disk\ clouds} \\
\sigma_{pc} &= 1.08\pm 0.01 \kms \quad\quad {\rm for\ Galactic\ center\ clouds} 
\end{split}
\label{eq:sigmapc}
\end{equation}
\noindent
From data in our Galaxy alone, one can thus already conclude 
that CBTCs vary in different environments within a galaxy.
A two-sample KS test between 
the $\sigma_{pc}$ distribution of 
the Galactic disk clouds and that of the Galactic center clouds 
gives a p-value $p=5\times 10^{-20}$, indicating they are statistically different.
It follows then that CBTCs and hence Larson's Relation may vary across galaxies
and in different environments within galaxies.
This prediction is supported by recent observations of molecular clouds in other galaxies
\citep[e.g.,][]{2013DonovanMeyer,2013Hughes,2014Colombo, 2020Krieger}.

Historically, from Eq (\ref{eq:grav}) we see that, if one insists expressing the Larson's first relation 
with the exponent close to $0.5$, 
the original Larson's first relation would be gas cloud surface density dependent,
a point later re-iterated \citep[][]{2009Heyer}.  
But if the range in $\Sigma_R$ is sufficiently narrow, one would obtain the original scale of a slope of $1/2$,
which may be the reason for that result obtained by \citet[][]{1981Larson}.
Thus, the original Larson's first relation has a limited scope and is applicable only
when the range of surface density is narrow enough.
In contrast, the revised Larson's Relation, Eq (\ref{eq:sigmaR2}),
is expected to be valid universally,
except that the strength parameter, $\sigma_{pc}$,
is expected to vary across different environments and across galaxies.
To illustrate this point better, let us express $\sigma_{pc}$ in terms of direct observables,
involving gas surface density.
Combining Eq (\ref{eq:grav}) and Eq (\ref{eq:sigmaR2}) gives
\begin{equation}
\begin{split}
\sigma_{pc} &= \alpha_v^{3/10} ({\Sigma_R\over 341 {\rm \msun~pc^{-2}}})^{1/2} ({R\over 1pc})^{-1/10}~\kms.
\end{split}
\label{eq:sigSig}
\end{equation}

The large difference between the Larson's Relation for the disk clouds and the Galactic center clouds
strongly indicates an important role played by turbulence and that
the CBTCs in the disk and at the Galactic centers are different, since gravity is the same.
While one may use Eq (\ref{eq:sigmaR2}) or Eq (\ref{eq:sigSig}) or other variants
to drive $\sigma_{pc}$ empirically with three observables,
such a derivation does not address the physical origin of the magnitude of $\sigma_{pc}$.
A simple top-down illustrative method to derive $\sigma_{pc}$ is given in \S 4.
We should note that our adoption of the Kolmogorov turbulence spectrum 
is largely motivated by available simulations.
The obtained consistency with observations suggests it may be valid.
The agreement with the observed fractal dimension in \S 3 is consistent with Kolmogorov spectrum as well.
Nonetheless, in general, the turbulence spectrum may not adhere strictly to that of Kolmogorov type.
A more general form of Eq (\ref{eq:sigmaR2}) may be written as
\begin{equation}
{\rm \sigma_R \propto \alpha_{v}^{1/5} R^{(3\phi+2)/5}}.
\label{eq:uLg}
\end{equation}
\noindent
For the Kolmogorov turbulence, we have $\phi=1/3$, which yields an exponent of $0.6$.
For \citet[][]{1948Burgers} turbulence, we have $\phi=1/2$, corresponding to an exponent $0.7$,
while the turbulence in a strong magnetic field may have a 
Iroshnikov-Kraichnan \citep[][]{1964Iroshnikov,1965Kraichnan} type 
with $\phi=1/4$, which would yield an exponent of $0.55$.
If one were to ascribe the difference in the exponent for the Galactic disk and Galactic center clouds
to physical differences in the respective turbulence, one possible exit is that the turbulence on the Galactic disk
is closer to that of Iroshnikov-Kraichnan type than the turbulence at the Galactic center.
This requires further work to clarify that is beyond the scope of this paper.
Nonetheless, none of different types of turbulence is expected to yield
the conventional exponent for the Larson's Relation of $0.5$.


Another point maybe worth noting is that 
there are clouds with $\alpha_v<1$, i.e., over-virialized clouds.
Obviously, these clouds seem unlikely to be evolutionary descendants of clouds that had $\alpha=1$ 
and subsequently endured some gravitational collapse.
If that were the case, 
it would imply a turbulence dissipation time significantly less than the free-fall time
of the system, inconsistent with simulations \citep[e.g.,][]{1998Stone}.
Therefore, we suspect that these low $\alpha_v$ systems are
a direct product of turbulence, 
clouds that have relatively low velocity dispersion for their gravitational strength and are probably transient,
due to the randomness of the turbulence.

\begin{figure}[h!]
\centering
\vskip -0.0cm
\resizebox{5.0in}{!}{\includegraphics[angle=0]{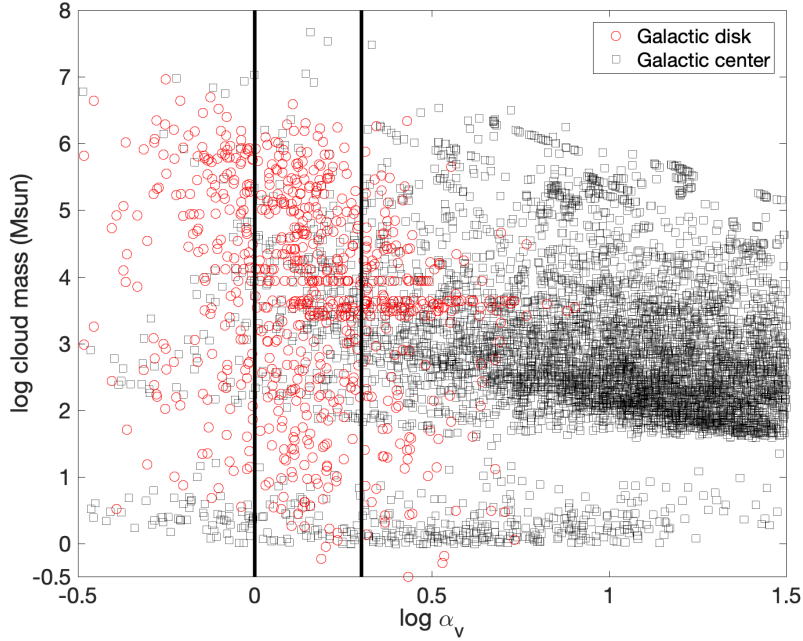}}
\vskip -0.5cm
\caption{
shows cloud mass as a function of $\alpha_v$ for 
the Galactic disk clouds (open red circle) and Galactic center clouds (open black squares).
The two vertical lines indicate clouds with $\alpha_v=1$ and $2$, respectively, for reference.
}
\label{fig:Malphav}
\end{figure}

To further clarify the nature of these special clouds,
we show in Figure (\ref{fig:Malphav})
the cloud mass as a function of its virial parameter.
To our surprise, clouds with $\alpha_v<1$ span the entire mass range.
This may be consistent with the randomness of the turbulence suggested above.
We note, however, that some of the most massive clouds ($\ge 10^6\msun$), i.e., giant molecular clouds,
may be a collection of uncommunicative, smaller clouds in an apparent contiguous region,
where the measured velocity dispersions reflect those of their smaller constituents,
while the overall gravitational energy increases with congregation;
we note that the velocity dispersion in this case may be significantly anisotropic.
Finally, the ubiquitous existence of gravitationally unbound clouds is simply due to insufficient
gravitational force relative to the turbulence velocity field in these clouds.
A point made here is that gravitationally unbound clouds are not necessarily those that
become gravitationally bound first and later become unbound due to internal stellar feedback or 
cloud-cloud collisions \citep[e.g.,][]{2011Dobbs}.
 
In Figure (\ref{fig:Malphav}) it is seen that
the clouds at the Galactic center (black squares) show a noticeable gap in mass, from $\sim 3\msun$ to $\sim 30\msun$.
It is not clear to us what might have caused this.
There is a separate ridge (horizontally oriented) of clouds near the bottom of the plot for the Galactic center clouds
with masses around one solar mass.
These low mass clouds appear to be mostly unbound.
While it is not definitive,
these clouds may be the counterpart of sub-solar mass clouds on the Galactic disk
called ``droplets" with odd ``virial" properties \citep{2019Chen,2019bChen},
although we are not sure why their typical mass is about $1\msun$ instead of $\sim 0.4\msun$ found for the droplets.
These small systems have large virial parameters but remain bound by 
external (thermal and turbulent) pressure.
The connection between these systems and the CBTC that we envision here
may no longer be direct,
and considerations of some additional physics may be required to place these systems also within 
the general framework outlined here.
We defer this to a later work.


Another word to further clarify the physical meaning of Eq (\ref{eq:sigmaR2}) may be in order,
which, let us recall, is a result 
derived based on the joint action of the statistical order imposed by turbulence of strength $\bar\sigma_{pc}$ (with a small 
dispersion)
and the natural selection effect by self-gravity,
with (the inverse of) $\alpha_{v}$ describing the strength of the latter acting against the former.
If $\alpha_{v}$ is much greater than unity, gravitational force would be too feeble to hold
the cloud together long enough to dissipate the excess energy to allow for further consistent gravitational contraction 
in the presence of internal and external disruptive force of turbulence.
Thus, the observed clouds with $\alpha_{v}$ greatly exceeding unity that are products of supersonic turbulence
are likely transient in nature.
Nonetheless, they may be useful for some physical analysis.
They may be considered good candidates for analyses where a statistical equilibrium is a useful assumption.
At the other end, when $\alpha_{v}$ is close to unity,
gravitational collapse of a cloud may ensue, detaching it from the parent CBTC. 
However, as noted in Figure (\ref{fig:Malphav}),
one should exercise caution to treat clouds with an apparent $\alpha_v$ less than unity
that may not be genuinely coherent gravitational entities, ready to run away and collapse.
We shall not delve into this further but note that these apparently over-virialized clouds
may not possess the usual gravitationally induced density stratification and may lack a
coherent structure (such as a well defined center).

\section{Fractal Dimension of the ISM}

Using Eq (\ref{eq:grav}) and Eq (\ref{eq:sigmaR2}), 
we may express the cloud density-size relation:
\begin{equation}
\begin{split}
n_R &= \alpha_v^{-3/5} {153\over 44\pi}{\sigma_{pc}^2\over G m_p (1pc)^2} ({R\over 1pc})^{-4/5} \\
&=1.0\times 10^4\alpha_{v}^{-3/5} ({\sigma_{pc}\over 1\kms})^2 ({R\over 1pc})^{-4/5}~cm^{-3},
\end{split}
\label{eq:nR}
\end{equation}
\noindent
where $n_R$ is the mean hydrogen number density within radius $R$ and $m_p$ is proton mass.
Then, the size-cloud mass relation follows:
\begin{equation}
\begin{split}
M_R =2.6\times 10^2\alpha_{v}^{-3/5} ({\sigma_{pc}\over 1\kms})^2 ({R\over 1pc})^{11/5} \msun.
\end{split}
\label{eq:ML}
\end{equation}
\noindent
Since $\alpha_v$ and $R$ are uncorrelated, for clouds generated by a same CBTC (a $\sigma_{pc}$ with dispersion),
we see that $M_R\propto R^{11/5}$.
\refp{\therefcounter}{
This expression usually has a slope of 2 when calculated based on dust extinction, a more accurate mass tracer than CO. 
(See extensive work by Lada/Alves/Lombardi on this, mentioned in Beaumont CN, Goodman AA, Alves JF, Lombardi M, Román-Zúñiga CG, Kauffmann J, Lada CJ (2012) A simple perspective on the mass-area relationship in molecular clouds. Monthly notices of the Royal Astronomical Society, 423:2579–2586. https://doi.org/10.1111/j.1365-2966.2012.21061.x https://doi-org.ezp-prod1.hul.harvard.edu/10.1111/j.1365-2966.2012.21061.x). The Beaumont et al. paper shows that the $M~R^2$ relation is also likely "fake," and due to the small range of column density PDF means.
}
\setcounter{tempcounterb}{\value{refcounter}}
\stepcounter{refcounter}
This mass-size relation with a slope of $2.2$ is in excellent agreement with observed value
of $2.2\pm 0.1$ \citep[][]{2001Heyer},
\stepcounter{refcounter}
and $2.36\pm 0.04$ \citep[][]{2010RomanDuval}.  

There are many different techniques used to measure cloud mass and size.
We stress that the size-mass relation depends on how clouds are defined or selected.
For the same reason that the original Larson's size-velocity dispersion relation has 
an exponent of $1/2$,
the original Larson's size-mass relation has an exponent of $2$.
Both are due to a small surface density range of the clouds \citep[e.g.,][]{2012Beaumont}.
The exponent in Eq (\ref{eq:ML}) expresses the size-mass relation for clouds at a fixed virial parameter.

In the context of a fractal, self-similar structure, which may approximate the ISM reasonably well,
Eq (\ref{eq:ML}) indicates that 
the fractal dimension of the ISM 
is $D=2.2$ \citep[][]{1983Mandelbrot} 
with the implied size function of the form 
\begin{equation}
n(L)dL \propto L^{-D-1}dL \propto L^{-16/5}dL.
\label{eq:nL}
\end{equation}
\noindent
The slope $16/5$ in Eq (\ref{eq:nL}) is 
in excellent agreement with the observed value of $3.2\pm 0.1$ for CO detected molecular clouds
in the Milky Way spanning the range of $\sim 1-100$pc \citep[][]{2001Heyer}.

The fractal dimension of the ISM of $D=2.2$ corresponds
to density power spectrum of $P_k\propto k^{D-3}\propto k^{-0.8}$.
It is helpful to have an intuitive visualization of this outcome.
In the process of energy transmitting downward along the spatial/mass scale via supersonic motion,
shocks and radiative cooling,
density structure (density fluctuation spectrum) is generated.
In three dimensional space,
an ideal, long and uniform filament will have a density power spectrum
$P_k\propto k^{-1}$ on scales below the length of the filament.
Similarly, a uniform sheet corresponds to $P_k\propto k^{-2}$,
whereas a point corresponding to a density power spectrum of $P_k=k^{0}$.
In absence of self-gravity,
compressive supersonic turbulence with sufficient cooling has the tendency to form filaments where
two planar shocks intersect.
In realistic situations with self-gravity, filaments have varying lengths and
the actual density power spectrum is expected to deviate somewhat from this, depending on the nature of
driving and energy distribution of the driving,
and the power spectrum is in general $P_k\propto k^{-\beta}$ with $\beta<1$.
Nevertheless, as long as the energy in the turbulence is dominated 
on the large scales, $\beta$ is not likely to be much less than unity. 
Thus, we see that the Kolmogorov compressive turbulence generated,
\refp{\therefcounter}{
More like Kolmogorov compressive turbulence "could be generated" not "is generated"... many slopes "not much less than unity" are possible with these arguments, right?
}
\stepcounter{refcounter}
gravitationally significant structures, in the presence of rapid radiative cooling,
have a density structure that is dominated by filamentary structures with a small mixture of knots.

\section{Estimate $\sigma_{pc}$ For Viscously Driven Turbulence}

In the normal situation where star formation occurs on a disk, 
it is reasonable to assume that the radius of the largest turbulence ``cloud",
which will be the driving scale of the CBTC, 
is equal to the scale height of the disk for isotropic turbulence.
This driving scale, $R_{d}$, can be expressed as
\begin{equation}
R_d= {C R_g\sigma_d^2(R_g)\over v_c^2(R_g)},
\label{eq:Rd}
\end{equation}
\noindent
where $\sigma_d(R_g)$ is the velocity dispersion on the driving scale $R_d$ at a galacto-centric radius $R_g$, 
which is also the vertical dispersion, $v_c(R_g)$ is the circular velocity at radius $R_g$,
and $C$ is a constant of order unity to absorb uncertainty.
We shall assume that the energy source is the rotational energy at the location,
where the turbulence may be driven by some viscous processes on the disk.
With such an assertion, one can relate $\sigma_d$ to $R_d$ by 
\begin{equation}
\sigma_d= 2 B R_d \Omega(R_g), 
\label{eq:sigmad} 
\end{equation} 
\noindent
where $\Omega(R_g)$ is the angular velocity at the radius $R_g$ 
for a Mestel disk that we will adopt as a reasonable approximation,
and $B$ is another constant of order unity to absorb uncertainty.
For a gas cloud (assumed to be uniform) of radius $R_d$, we can express the virial parameter by
\begin{equation}
\alpha_d= {3\sigma_d^2(R_g) \over {3\over 5}{GM_d\over R_d}} = {15\sigma_d^2\over 4\pi G\rho_d R_d^2},
\label{eq:alphad}
\end{equation}
\noindent
where $\rho_d$ is the gas density at the driving scale.
With Eq (\ref{eq:Rd},\ref{eq:sigmad},\ref{eq:alphad}) we can compute
$\sigma_{pc}$ using Eq (\ref{eq:sigmaR2});
\begin{equation}
\sigma_{pc}= 0.44\kms ({D\over 2.3})^{-1/5} ({\Sigma_d\over 5\msun~{\rm pc}^{-2}})^{1/5} ({v_c\over 220\kms})^{3/5} ({R_g\over 8{\rm kpc}})^{-2/5},
\label{eq:alphad3}
\end{equation}
\noindent
where we have defined another constant $D\equiv B/C$.
Eq (\ref{eq:alphad3}) is expressed such that
if the fiducial values are taken, we obtain $\sigma_{pc}=0.44\kms$ for disk clouds center near the solar radius,
as derived earlier (see Eq \ref{eq:sigmapc}).
Aside from the unknown combination of $D$, all other fiducial values are well observed,
including the gas surface density of $5\msun~{\rm pc}^{-2}$ \citep[e.g.,][]{2017Sofue}.
Interestingly, if we use the same $D=2.9$ value along with the relevant values for other parameters
for the Galactic center, 
$\Sigma_d=30\msun~{\rm pc}^{-2}$ \citep[][]{2017Sofue},
$R_g=500$pc (within which the Galactic center clouds are observed), $v_c=250\kms$ \citep[][]{2017Sofue},
we obtain $\sigma_{pc}= 2.1\kms$, larger than the value of $1.08\kms\pm 0.01$dex, derived
for the clouds at the Galactice center (see Eq \ref{eq:sigmapc}).
Although the expectation that $\sigma_{pc}$ at the Galactic center is larger than
that on the Galactic disk is in agreement with the derived values,
the numerical discrepancy may be due to a number of causes.
It may be in part due to different observational systematics for disk clouds and center clouds.
It may be in part due to that the treatment of the central region of the Galaxy as a disk breaks down
or that the effective viscosity in the two regions are different.
It is notable that our simple calculations 
do not require participation of some other physical processes
that might be relevant, including magnetic field, stellar feedback.
While this is not a vigorous proof of the veracity of our assumptions, 
the found agreement between the predicted $\sigma_{pc}$ and the directly calculated value for
the Galactic center clouds is a validation of our basic assumptions and the resulting outcomes,
that is, turbulence and gravity play a dominant role in shaping the interstellar medium
and the formation of clouds down to at least the sonic scale.

\section{Conclusions}

An analysis of a joint action of compressive turbulence and self-gravity is performed.
Physically, it may be considered that the turbulence is bookended 
by the gravitationally significant clouds at the small scales, 
as opposed to the driving scale on large scale.
We denote such a turbulence chain as ``cloud bound turbulence chain" (CBTC).

The (new) Larson's Relation, $\sigma_R=\alpha_v^{1/5}\sigma_{pc} (R/1{\rm pc})^{3/5}$, 
relating the velocity dispersion $\sigma_R$ to the size $R$ of a cloud, is derived,
where $\alpha_v$ is the virial parameter of the cloud
and $\sigma_{pc}$, the velocity dispersion of the turbulence at $1$~pc, encodes the strength of the CBTC. 
Although implicit in the assumption is that the turbulence is supersonic,
the new Larson's Relation is shown to hold at least down to the transonic scale of $0.05$pc.
The conventional exponent of $1/2$ for the Larson's Relation is shown to be excluded.

The most significant finding is not necessarily the derivation of this relation naturally
and the exponent $3/5$ being in good agreement with observations.
It is prudent to remind ourselves that this exponent depends on the assumed Kolmogorov spectral index
for the turbulence according and error treatments of cloud measurements can certainly be improved
that may change its value to some extent. 
Rather, it is the fact, which is made plain by the analysis as well as empirical evidence,
that, while the exponent of the Larson's Relation may be universal or close to universal, the amplititude, $\sigma_{pc}$, 
is not and may differ greatly.
The latter is environment dependent, reflecting the dependence of $\sigma_{pc}$ of a CBTC on environment.
The recognition of and evidence for the non-universality of the Larson's Relation is of foundamental physical importance.
The implications may be profound for star formation process, 
which is thought to be dependent on the Mach number of turbulence,
which in turn is linearly proportional to $\sigma_{pc}$.

Our analysis also yields a by-product with respect to some properties of the fractal nature of the ISM.
We show that the fractal dimension of the ISM is $11/5$,
cloud (linear) size function of $n(R)dR\propto R^{_16/5}dR$,
both in nearly exact agreement with observations.

\vskip 1cm
I would like to thank the referee Dr. Alyssa Goodman for a constructive report that 
helped significantly improve the paper. 
I thank Dr. Mark Heyer for kindly providing the observational data in suitable formats,
Dr. Eric Koch for kindly sharing the CHIMPS2 survey data, along with Drs. Erik Rosolowsky, David Eden and Nico Krieger,
Dr. Frederic Schuller for kindly sharing the SEDIGISM survey data,
Dr. Nico Krieger for his kind help with obtaining data,
and many colleagues for useful discussions.
This work is supported in part by grant NASA NNX11AI23G.


\end{document}